%% file: cdc19poisson.tex
\documentclass[11 pt, onecolumn]{article}
\usepackage[margin=1.1in]{geometry}
\input{./sections/header}
\usepackage{graphicx}
\usepackage{epstopdf}

\begin{document}
\title{\LARGE \bf Evaluation of Headway Threshold-based Coordinated Platooning over a Cascade of Highway Junctions}

\author{Xi Xiong, Teze Wang, and Li Jin
\thanks{This work was in part supported by the NYU Tandon School of Engineering and the C2SMART University Transportation Center. The authors appreciate the discussion with Prof. Karl H. Johansson at the KTH Royal Institute of Technology.}% <-this % stops a space
\thanks{X. Xiong, T. Wang, and L. Jin are with the Department of Civil and Urban Engineering, New York University Tandon School of Engineering, Brooklyn, NY, USA, emails: xi.xiong@nyu.edu, teze.wang@nyu.edu, lijin@nyu.edu.}%
}
\newcommand*{\QEDA}{\hfill\ensuremath{\blacksquare}}%

\maketitle
%=======================================================================================================================
%==================================================

\begin{abstract}
Platooning of vehicles with coordinated adaptive cruise control (CACC) capabilities is a promising technology with a strong potential for fuel savings and congestion mitigation. Although some researchers have studied the vehicle-level fuel savings of platooning, few have considered the system-level benefits. This paper evaluates vehicle platooning as a fuel-reduction method and propose a hierarchical control system.
We particularly focus on the impact of platooning coordination algorithm on system-wide benefits.
The main task of platooning coordination is to regulate the times at which multiple vehicles arrive at a particular junction: these vehicles can platoon only if they meet (i.e. arrive within a common time interval) at the junction.
We use a micro-simulation model to evaluate a class of threshold-based coordination strategies and derive insights about the trade-off between the fuel savings due to air drag reduction in platoons and the extra fuel consumption due to the coordination (i.e. acceleration of some vehicles to catch up with the leading ones).
The model is calibrated using real traffic data of a section of Interstate 210 in the Los Angeles metropolitan area.
We study the relation between key decision variables, including the platooning threshold and the coordination radius, and key performance metric, fuel consumption.
\end{abstract}

%\begin{keywords}
{\bf Index terms}: Connected and autonomous vehicles, Vehicle platooning, Micro-simulation, Fuel savings.
%\end{keywords}

\input{./sections/introduction} %1p
\input{./sections/modeling} %1p
\input{./sections/evaluation} %3p
\input{./sections/background} %1p
\input{./sections/conclusion} %0.5p

\bibliographystyle{IEEEtran}
\bibliography{bib_LJ}   
\end{document}

%% file: sections/header.tex
\usepackage{amsmath}
\usepackage{bm}	
\usepackage{graphicx}

\usepackage{graphicx}
\usepackage{multirow}
\usepackage{mdwlist}

\usepackage{threeparttable}

\usepackage{amsfonts}
\usepackage{amsmath}
\usepackage{hyperref}
\usepackage{bbm}
\usepackage{float}
\usepackage{caption}
\usepackage{subcaption}
\usepackage{xcolor}
\usepackage{mathtools}
\usepackage{tikz}

\usetikzlibrary{shapes.geometric,arrows}
\usepackage{enumerate}
\usepackage{multicol}
\usepackage{stackrel}

\usepackage{subcaption}
\usepackage{amssymb}
\usepackage[utf8]{inputenc}
\usepackage{booktabs,lipsum}
\usepackage{hyperref}

%% file: sections/introduction.tex
\section{Introduction}

%background and challenge
Connected and autonomous vehicles (CAV) recently attract much attention due to the advancement in computation and communication technologies. The CAV technology enables vehicle platooning, a novel operation that can yield remarkable efficiency and environment benefits \cite{hor+var00,bess+16procieee,talebpour16,tsugawa16,litman2017autonomous}. Generally, we can divide the platooning process into two stages: (i) coordinating the arrival time of multiple vehicles at a junction and (ii) merging the vehicles in a platoon at the junction \cite{varaiya1993smart,larson2015distributed,xiong2019analysis}. Although significant progress has been made recently on the implementation of vehicle-level control to improve the safety and stability of forming platoons \cite{naus+10,coogan15interconnected,gao2017data,besselink2017string,hu2018plug}, we still lack systematic understanding of coordinating strategies for platooning over large-scale road networks.

%literature review
Previous work on CAV coordination for platooning emphasizes on the scheduling algorithms to coordinate the movement of CAVs. Importantly, Larson et al. \cite{larson2015distributed} proposed a network of distributed platooning coordinators that manage formation of platoons at highway junctions. Van de Hoef et al. \cite{van2018fuel} considered the en-route formation of platoons. In a related work \cite{xiong2019analysis}, we considered a simplified stochastic model for coordinated platooning at a single junction. Another line of work \cite{jin2018modeling,cicic2019multiclass,cicic2019coordinating} considers the coordination of CAV platoons in the presence of background traffic, but these results do not address the formation of platoons. To have a better insight of the impact of different coordinated platooning strategies, we need a system level framework to capture the interaction of CAVs over road networks with multiple junctions.

%our design of two layer system
In this paper, we consider a hierarchical control framework for coordinated platooning on a highway with a cascade of junctions and evaluate it using a micro-simulation testbed. The framework is a two-layer system. The upper layer coordinates the movement of vehicles to change the arrival time to form a platoon. We mainly focus on this layer, which is the essential factor to determine the probability of forming platoons. The lower layer is the merging process including longitude and latitude control in vehicle level; this layer is involved in our micro-simulation experiments, but we do not explicitly study this layer in detail.

\begin{figure}
  \centering
  \includegraphics[width=0.65\textwidth]{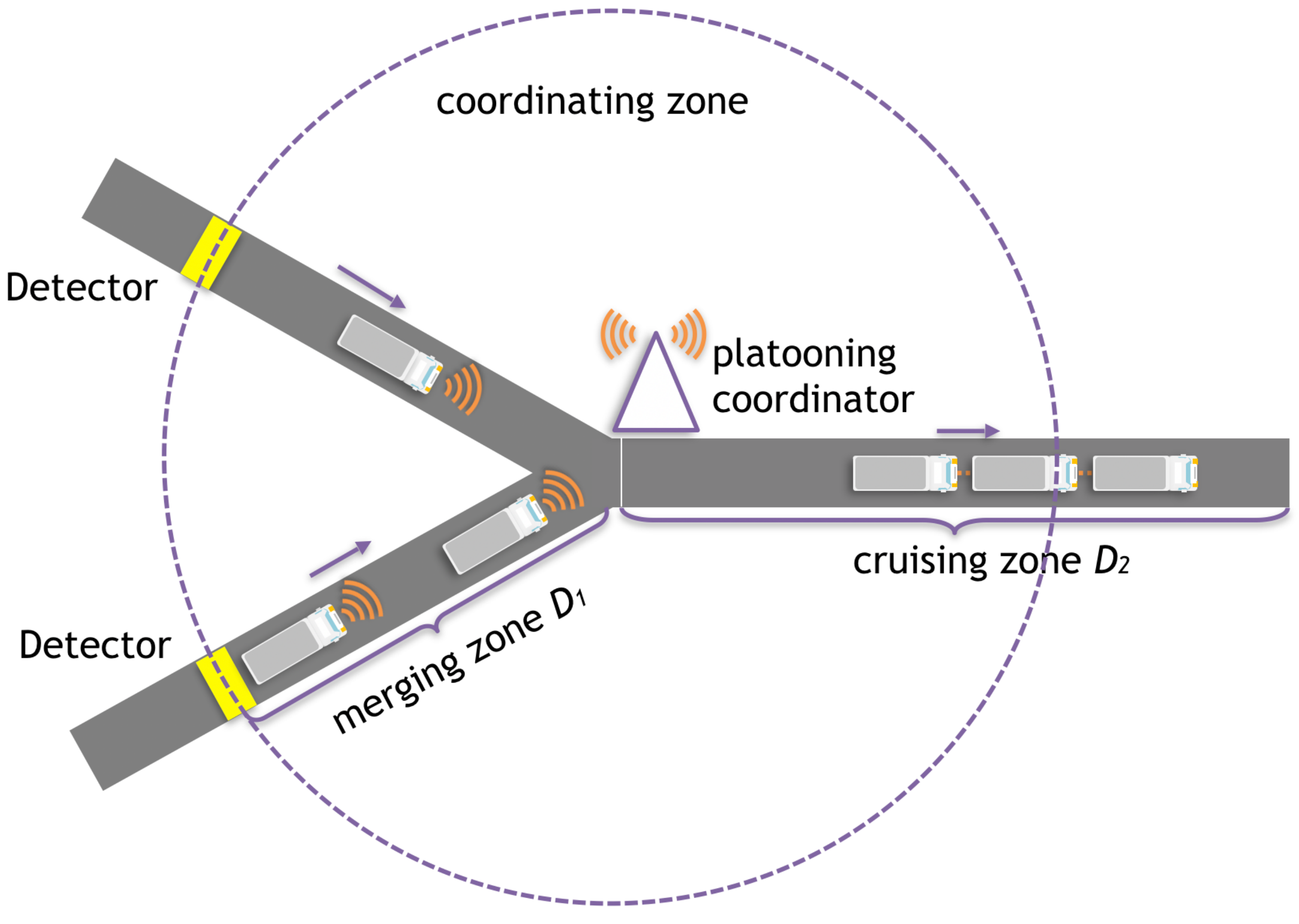}
  \caption{Coordinated platooning at a highway junction.}\label{fig:color_figure}
\end{figure}

Figure \ref{fig:color_figure} illustrates the scenario we consider in this paper. In the upper level control, two detectors are placed on both branches before the junction. Once a vehicle passes a detector, the platooning coordinator estimates the time at which this vehicle will arrive at the junction and determines the platooning policy based on the arrival times. We focus on a simple but practically relevant threshold-based strategy to form platoons: the following vehicle would be instructed to merge with the leading vehicle if the headway (i.e. the difference between the estimated times of arrival at the junction) between the two vehicles is below a given threshold. The following vehicle would adapt the speed to catch up with leading vehicle when two vehicles are assigned to form a platoon. The leading vehicle and following vehicle are coordinated to arrive at the junction almost at the same time, which would facilitate the process of merging. In our approach, we can use the two-layer system to implement coordinated control using real traffic data.

In the process of merging, the following vehicle would accommodate the speed and lateral movement to follow the leading vehicle. Generally, the movement of coordination includes following vehicle acceleration, leading vehicle deceleration, and a mixture of acceleration and deceleration. In this paper, we follow \cite{larson2015distributed} and only consider the case of  acceleration; i.e. the following vehicle would catch with the leading vehicle. Since the following vehicle arrives not far from the leading vehicle, the merging process would hardly affect the vehicle behaviors. After the junction, the formed platoon would enter the cruising zone on the highway. 

The objective of forming platoons is to improve traffic efficiency and environmental benefits. In this paper, we mainly focus on the fuel savings thanks to platooning. In the coordinating zone before the junction, the following vehicle would increase the speed to catch up with the leading vehicle, which would increase the fuel consumption. While in the cruising zone after the junction, vehicles in a platoon would experience reduced aerodynamic drag and improve fuel economy, which would decrease fuel consumption. In this paper, we focus on analyzing the trade-off between fuel consumption during the coordinating phase and during the cruising phase.
We conduct our analysis over a testbed calibrated for a section of the Interstate 210 in the Los Angeles area in Simulation of Urban MObility (SUMO), a standard micro-simulation tool \cite{krajzewicz2002sumo}. The traffic flow data are collected from the Caltrans Performance Measurement System (PeMS) \cite{varaiya2007freeway}. Total fuel consumption is calculated to analyze different parameters in the proposed structure. 

%evaluation with background and without background
We first consider threshold-based control at single junction; of particular interest is the impact of the threshold on total fuel consumption. Our results imply that when the cruising distance after the junction is small, the effect of acceleration before the junction weights over the effect of reduced fuel consumption over the cruising distance. Total fuel consumption would increase as we increase the threshold, which represents the chance of forming platoons. When the cruising distance is long enough, reduced fuel economy due to platooning would play the major part.
We also consider the fuel savings under various levels of CAV flow. It is intuitive that a larger flow of CAVs leads to more chances for platooning and thus more fuel savings; our analysis quantifies this relation.

Then, we study coordinated platooning over multiple junctions. Since cruising distance is long enough in this scenario, total fuel consumption would decrease as we increase the threshold. We also investigate the effect of different placement of detectors. Our results indicate that longer distance of detectors would decrease fuel consumption. We also reveal that adding more connected vehicles would increase fuel economy due to the increased chance of forming platoons. In addition, we investigate the coordination of two junctions in the network. The results show that the minimum value is not obtained at the largest threshold in both junctions, which indicates that the interaction of traffic demands in the network would affect the final fuel consumption. Furthermore, we study the effect of background vehicles on total fuel consumption \cite{jin2018modeling}. The results show that adding more background vehicles would change the behaviors of connected vehicles.

%our contributions
The main contributions of this paper include: (i) development and implementation of a hierarchical control system for coordinated platooning in a practically relevant setting; (ii) analysis of the relation between fuel consumption and key operation parameters, including the headway threshold for platooning, the incoming rate (flow) of CAVs, the road geometry, and the level of background traffic; (iii) validation of the theoretical results in a related work \cite{xiong2019analysis} derived from a simplified model in the more realistic micro-simulation environment.

%outline
The rest of the paper is organized as follows.
In Section~\ref{section:model}, we formulate the problem and propose the hierarchical system.
In Section~\ref{section:without_background_traffic}, we analyze the characteristics of the coordinated system without background vehicles.
In Section~\ref{section:with_background_traffic}, we analyze scenario with background vehicles.
In Section~\ref{section:conclusion}, we summarize the conclusions and discuss future work.

%% file: sections/modeling.tex
\section{Modeling coordinated platooning} \label{section:model}
In this section, we model the coordinated platooning process in a highway network. We first introduce the concept of hierarchical control in networks. Then we present the details of platooning process at a cascade of junctions. Finally we present our objective function: minimizing total fuel consumption.

\subsection{Hierarchical Control System}
Consider a given transportation network with $N$ junctions (including on-ramps and off-ramps), we propose the concept of hierarchical control for vehicle operation when coordinated platooning in the network. The upper level is merging control, and the lower level is vehicular control of platooning process. What we focus on is the upper level control since it determines the probability of forming platoons in the lower level. The lower level control is actually the vehicular platooning process including lateral and longitudinal control to form a platoon, which generates the benefit of fuel reduction (Figure \ref{fig:color_figure}).

In the upper level control, a system operator can use sensors (loop detectors) to receive the arrival time of each vehicle, which can be used to accommodate the vehicle motion to decide whether to form a platoon or not. Using this method, we can coordinate the platooning process at junctions. Given the sensor network in the highway, we can coordinate the platooning in a large scale.

We first consider the coordination at one junction of the mainline and an on-ramp, and assume the system operator can detect the arrival of incoming vehicles at a radius of $D_1$. Vehicle on the mainline and off-ramp should drive $D_1$ before entering the junction. The coordination process includes two steps: determining which vehicle to platoon and planning the vehicle motions. The first step is the principle of platooning using the arrival time from the two sensors in the mainline and on-ramp. In the following, we propose a threshold-based principle, which is simplistic but non-trivial for implementation. The second step during platooning is the vehicular motion coordination. Generally, the vehicular coordination is flexible in merging process: leading vehicles can be accelerated, following vehicles can be decelerated, or the mixture of both methods. Since acceleration and deceleration are symmetrical, to simplify the modeling process, we only consider the acceleration option for following vehicles. Our approach can be extended to deceleration. 

In the lower level control, we need to coordinate both the leading and following vehicle, which also includes two steps: catching up  and platooning. Two distances are related to theses two stages: $r_1$ and $r_2$, in which we assume $r_2 \leq r_1$. If the distance between leading and following vehicle is less than $r_1$, the following vehicle enters the catching up stage, and the acceleration increases to catch the leading vehicle while the leading vehicle remains the same speed. When the distance between these two vehicles reaches $r_2$, they form the platoon at the first time. Since the following vehicle speed is still high at the time, the two vehicles need to accommodate respective speeds to remain the $r_2$ gap.

In our hierarchical control, if two vehicles are determined to form a platoon, the following vehicle would drive to meet the leading vehicle in a small time interval, which can make the lower level control more convenient. After merging at the junction, the following vehicle in a platoon would experience reduced aerodynamic drag and therefore improve fuel economy.

\subsection{Upper Level Coordination}
In this part, we analyze the platooning process in the upper level control in a mathematical perspective, which also affects the lower-level platooning forming process. We design the coordination system that controls the following vehicle which passes through the detectors to change speed to arrive at junction at the same time of the leading vehicle, in which the control principle is given based on the arrival time on each detector. In this paper, we do not explicitly involve lower-level controllers.

Consider the scenario where the detectors are placed on the mainline and on-ramp, and the distance is $D_1$ before the junction. We assume the average vehicle speed is $V_0$, then the average traverse time $T_0 =  D_1 / V_0$. Each vehicle's arrival time on the mainline and on-ramp could be recorded through corresponding detector. The arrival times on the mainline are denoted by the array $[t_0^0, t_0^1, t_0^2, \cdots, t_0^k]$, and the arrival times on the on-ramp are given by $[t_1^0, t_1^1, t_1^2, \cdots, t_1^m]$, where $k $ and $m$ are numbers of counts by each detector.

We define the leading vehicle as the vehicle that firstly passes either detector and the following vehicle as the vehicle that also passes either detector consequently. That is to say the vehicle pair could come from the same road branch or different ones, which is only differentiated by respective arrival time. We assume the leading vehicle operates on the mainline with speed $V_l$, and the arrival time at the detector on the mainline is $t_0^l$. The following vehicle cruises on the on-ramp towards the junction at the speed $V_f$, and the arrival time at the detector is $t_0^f$. Then the assumed leading vehicle arrival time at the junction is:
\begin{align} \label{Eq.leading_arrival_time}
     t_0^j = t_0^l + D_1 / V_0.   
\end{align}
We use the arrival time $t_0^l$ and $t_0^f$ to determine whether two vehicles to form a platoon. The principle is given by a simple but realistic threshold-based control as follows:
\begin{align}
   f(r) = \left\{
    \begin{array}{rcl}
    0 & &{t_0^f - t_0^l \geq r}\\
    1 & &{t_0^f - t_0^l < r}
    \end{array} \right.,  
\end{align}
where $f(r)$ represents the binary decision of whether to form a platoon: $f(r)=1$ means that the following vehicle would accelerate to meet the leading vehicle at the junction, and $f(r)=0$ means not to merge.

If $f(r) = 1$, the following vehicle is supposed to enter the junction at $t_1^j$, where $t_1^j = t_0^j$. Then the following vehicle's average speed $V_f$ should satisfy the following equation:
\begin{align} \label{Eq.following_arrival_time}
    t_1^j = t_0^f + D_1 / V_f.
\end{align}
We use Equation \ref{Eq.leading_arrival_time} and Equation \ref{Eq.following_arrival_time} to derive the average vehicle speed of the following vehicle as:
\begin{align*}
    V_f = \frac{D_1}{D_1 / V_0 - (t_0^f - t_0^l)}.
\end{align*}
To form a platoon with the leading vehicle, the following vehicle would change its speed to $V_f$. In practice, the leading vehicle would also be in the process of acceleration, which would result in the following vehicle with higher target speed. We use a speed limit $\hat{V_f}$ to bound the catching up process in Equation \ref{Eq.speed_limit}. The following vehicle would continue to catch up with speed $\hat{V_f}$ if $V_f > \hat{V_f}$, in which the following vehicle would not meet with the leading vehicle.
\begin{align}\label{Eq.speed_limit}
    V_f \leq \hat{V_f}.
\end{align}{}
Notice that the leading vehicle may cruise over the junction while the following vehicle arrives at the detector if we use a large threshold. We would not change the following vehicle behavior in this case.

\subsection{Minimizing Fuel Consumption}
The objective of coordinated platooning is to minimize fuel consumption. In the SUMO simulation, the function of fuel rate (fuel consumption per second, $mL/s$) is given as follow:
\begin{align}\label{Eq.fuel_rate}
   f = c_0 + c_1 v a + c_2 v a^2 + c_3 v + c_4 v^2 + c_5 v^3,
\end{align}
where $c_i, i=0,1,\ldots,5$ are constant parameters, $a$ is the acceleration, and $v$ is the speed. When the vehicle moves in the $D_1$ zone, we assume the variation of speed (acceleration) is small. Then Equation \ref{Eq.fuel_rate} can be simplified as
\begin{align*}
   f = c_0 + c_3 v + c_4 v^2 + c_5 v^3,
\end{align*} 
Consider a vehicle $k$ entering the junction, the driving distance after the junction is $D_2$. Then the fuel consumption can be divided into two parts: cruising on $D_1$ and cruising on $D_2$. We use $F^k_1$ to denote the fuel consumption in $D_1$ and use $F^k_2$ to represent the fuel consumption in $D_2$. In our case, we value the incremental total cost over the absolute value. Then the incremental total fuel consumption $\Delta{TC_k}$ for vehicle $k$ can be expressed as:
\begin{align*}
   \Delta{TC_k} = \Delta{F^k_1} + \Delta{F^k_2},
\end{align*}
where $\Delta{F^k_1}$ and $\Delta{F^k_2}$ represent incremental fuel consumption in $D_1$ and $D_2$.

If vehicle $k$ is set to catch up with leading vehicle, the first increased fuel consumption is:
\begin{align*}
   \Delta{F^k_1} & = s^k_f \left( c_3 V_f + c_4 {V_f}^2 + c_5 {V_f}^3 \right) - s^k_0 \left( c_3 V_0 + c_4 {V_0}^2 + c_5 {V_0}^3 \right) \\
               & = s^k_f \left[ c_3 \frac{D_1}{D_1 / V_0 - (t_0^f - t_0^l)} + c_4 \left( \frac{D_1}{D_1 / V_0 - (t_0^f - t_0^l)} \right)^2 + c_5 \left( \frac{D_1}{D_1 / V_0 - (t_0^f - t_0^l)} \right)^3 \right]  \\
               & \quad - s^k_0  \left( c_3 V_0 + c_4 {V_0}^2 + c_5 {V_0}^3 \right),
\end{align*}
where $s^k_f$ is the traverse time over $D_1$ using target speed $V_f$, and $s^k_0$ is the original traverse time using speed $V_0$.

After the junction, the following vehicle merges with the leading vehicle. Since the following vehicle in a platoon experience reduced aerodynamic drag and improve fuel economy, the second incremental total cost $\Delta{F^k_2}$ is negative. Trailing vehicles in a platoon save a fraction $\eta \in (0,1)$ of fuel used when driving alone, which is typically $0.05 \sim 0.15$ \cite{larson2015distributed}. We use $\theta$ to represent fuel efficiency (a ratio of distance traveled per unit of fuel consumed, $L/km$), then the second fuel consumption can be expressed as:
\begin{align*}
   \Delta{F^k_2} = -\eta \theta D_2.
\end{align*}
Then the incremental total fuel consumption can be expressed as:
\begin{equation}
\begin{aligned} \label{Eq:final_fuel_consumption}
   \Delta{TC_k} & = s^k_f \left[ c_3 \frac{D_1}{D_1 / V_0 - (t_0^f - t_0^l)} + c_4 \left( \frac{D_1}{D_1 / V_0 - (t_0^f - t_0^l)} \right)^2 + c_5 \left( \frac{D_1}{D_1 / V_0 - (t_0^f - t_0^l)} \right)^3 \right]   \\
               & \quad - s^k_0  \left( c_3 V_0 + c_4 {V_0}^2 + c_5 {V_0}^3 \right) -\eta \theta D_2.
\end{aligned}
\end{equation}
From the equation, we can see that $\Delta{F^k_1}$ increases the total fuel consumption due to acceleration, and $\Delta{F^k_2}$ decreases total fuel consumption due to reduced fuel economy in a platoon, which can be formulated as an optimization problem.

%% file: sections/evaluation.tex
\section{Evaluation and Design Without Background Traffic} \label{section:without_background_traffic}

In this section, we use experiment results in SUMO to show the performance of threshold-based coordinated platooning under various scenarios. We here focus on the scenarios where only connected and autonomous vehicles cruise on the highway; no other traffic is sharing the highway. We first illustrate the setting and dataset, then we present the threshold-based control at one junction, and finally present homogeneous and hetergeneous control at multiple junctions.

\subsection{Setting and Dataset}
\begin{figure}[!ht]
  \centering
  \includegraphics[width=0.9\textwidth]{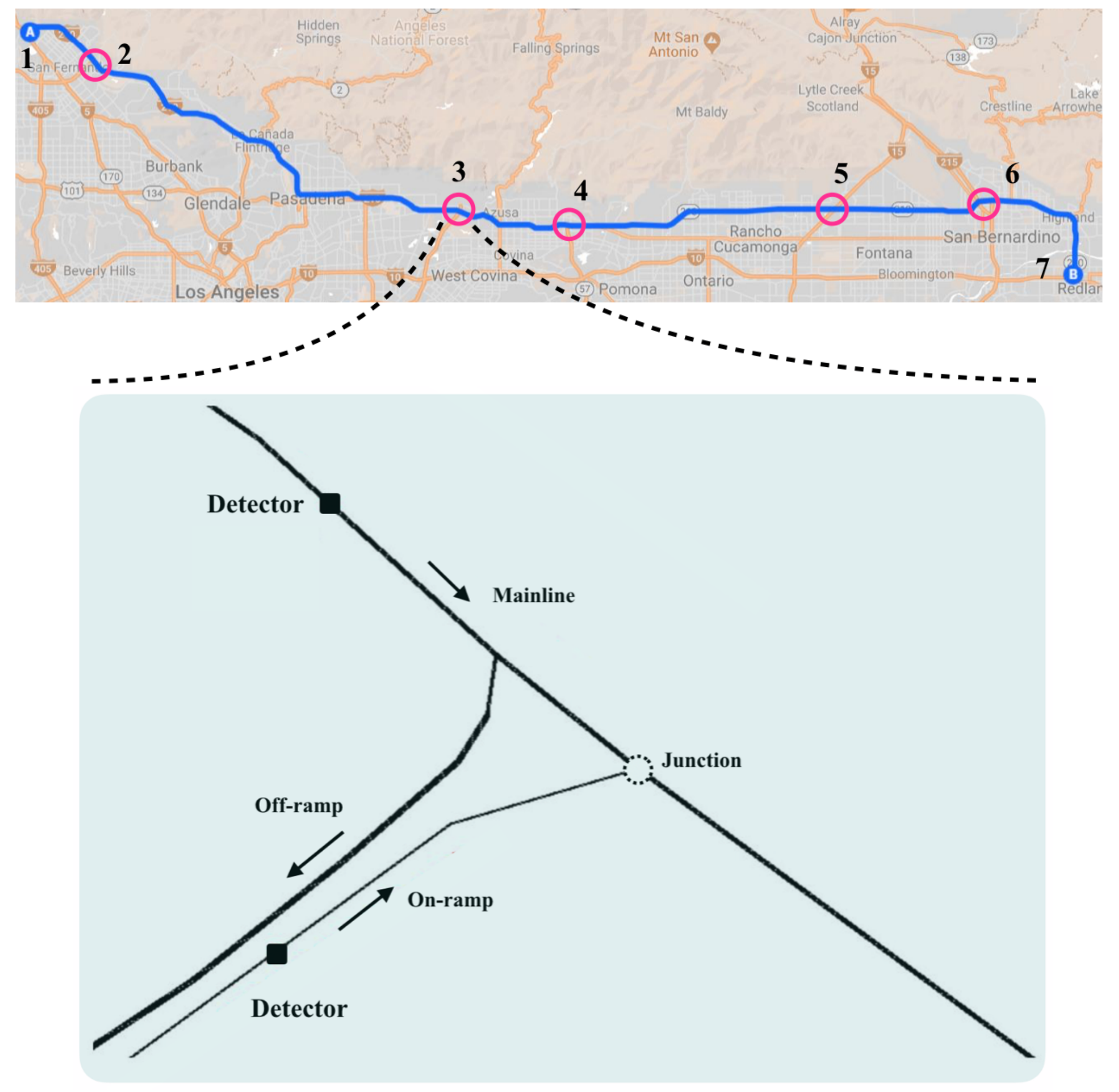}
    \caption{Network in SUMO}\label{fig:networks}
\end{figure}

We chose a highway section of I-210 in Los Angeles metropolitan to conduct our simulation (Figure \ref{fig:networks} top). We constructed the highway networks in SUMO and selected 5 junctions including on-ramps and off-ramps to implement connected vehicles. Figure \ref{fig:networks} bottom shows the network structure in the simulation platform. To simplify the implementation, we only consider the eastbound flow on the highway. In Figure \ref{fig:networks}, we consider 7 nodes (5 junctions, start node, and end node). Platoons are formed at each junction due to the coordination on the mainline and the on-ramp.

Detectors are placed $D_1$ before the junction on the mainline and the on-ramp. The passing time of each vehicle is recorded using two detectors, which can be used to determine the behavior of the following vehicle. If the time arrival interval is smaller than the given threshold $r$, the following vehicle would catch up with the leading vehicle. In the scenario where the formed platoon would include many vehicles, the following vehicle at the end of the platoon would accelerate at higher speed. We limit the vehicle speed to $40$ mph to reduce the effect of endless acceleration. Notice that the leading vehicle or the following vehicle can also leave the I-210 through the off-ramp, in which the platoon would not be formed.

We use the traffic data in Caltrans Performance Measurement System (PeMS) for simulation \cite{varaiya2007freeway}. The chosen analysis period is the peak hour from 8:00 A.M. to 11:00 A.M. on Tuesday, January 22, 2019. Since PeMS data only provide traffic counts at link detectors, we assume the ratio from the origin to the destination is proportion to the ratio of off-ramp vehicles at each junction. Then we can present the O-D matrix in 3 hours in Table \ref{table:dataset}. Each row denotes the flows from one origin to all other destinations. Since we only consider the flows in eastbound, there are only 21 O-D pairs in each time interval. We use the O-D matrix to generate flows in the network and calculate the total fuel consumption for all vehicles.

\begin{table}[!ht]
  \begin{center}
     \caption{Origin-Destination matrix in the highway of I-210 [veh/hr]}\label{table:dataset}
    \begin{tabular}{llllllll}
    \hline
    &                             & 2         & 3      & 4       & 5      & 6       & 7                                              \\
    \hline
    & (a) 8:00 - 9:00 A.M.                                                                                                  \\
    &1                                 &3040       &1240    &380      &460      &80       &1180                                            \\
    &2                               &-             &380     &340      &380      &160       &300                                             \\
    &3                               &- &-                      &100       &340 &120 &540                                                \\
    &4                    &- &- &-    &100 &120 &460                                            \\
    &5                   &- &- &- &-    &40 &420                                             \\
    &6                &- &- &- &- &-  &140                                             \\
    &7               &- &- &- &- &- &-                    \\
    \hline
    & (b) 9:00 - 10:00 A.M. \\
    &1                        &2140 &900 &480 &380 &60 &700                                             \\
    &2          &- &560 &300 &440 &100 &180                                              \\
    &3          &- &- &180 &280 &120 &380                                              \\
    &4         &- &- &- &180 &80 &440                                              \\
    &5         &- &- &- &- &40 &260                                              \\    
    &6         &- &- &- &- &- &160                                               \\
    &7      &- &- &- &- &- &-                   \\    
    \hline
    & (c) 10:00 - 11:00 A.M.  \\
    &1                &1360 &640 &500 &440 &60 &780                                              \\
    &2          &- &720 &200 &520 &100 &180                                              \\
    &3          &- &- &240 &220 &100 &340                                              \\
    &4          &- &- &- &240 &40 &420                                              \\
    &5          &- &- &- &- &60 &180                                             \\    
    &6          &- &- &- &- &- &200                                             \\
    &7          &- &- &- &- &- &-                         \\    
    \hline
    \end{tabular}
  \end{center}
\end{table}

\subsection{Threshold-based Control at One Junction}
\begin{figure}[!hbt]
  \centering
  %\subfigure{
  {\includegraphics[width=0.5\textwidth]{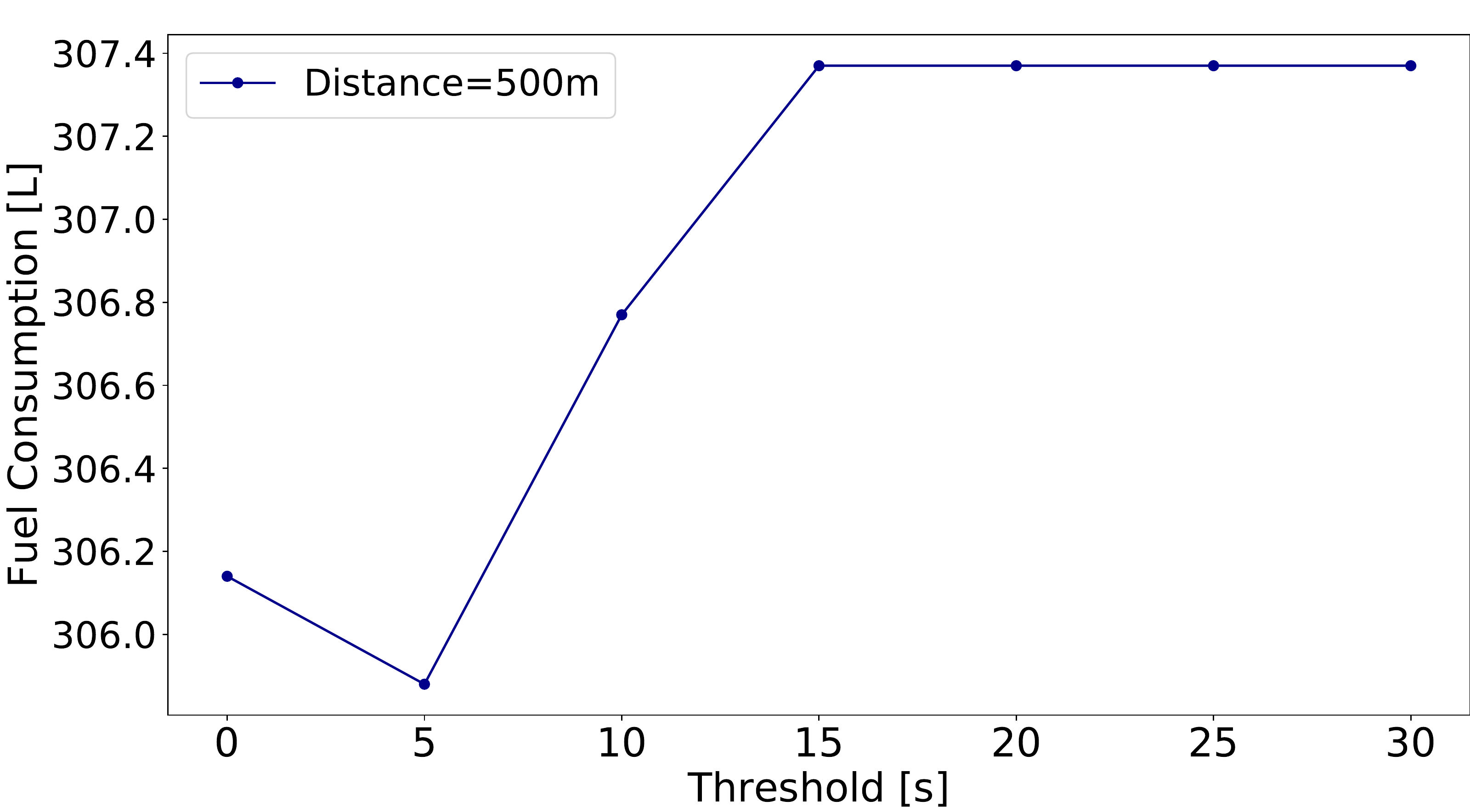}
  }

  %\subfigure
  {\includegraphics[width=0.5\textwidth]{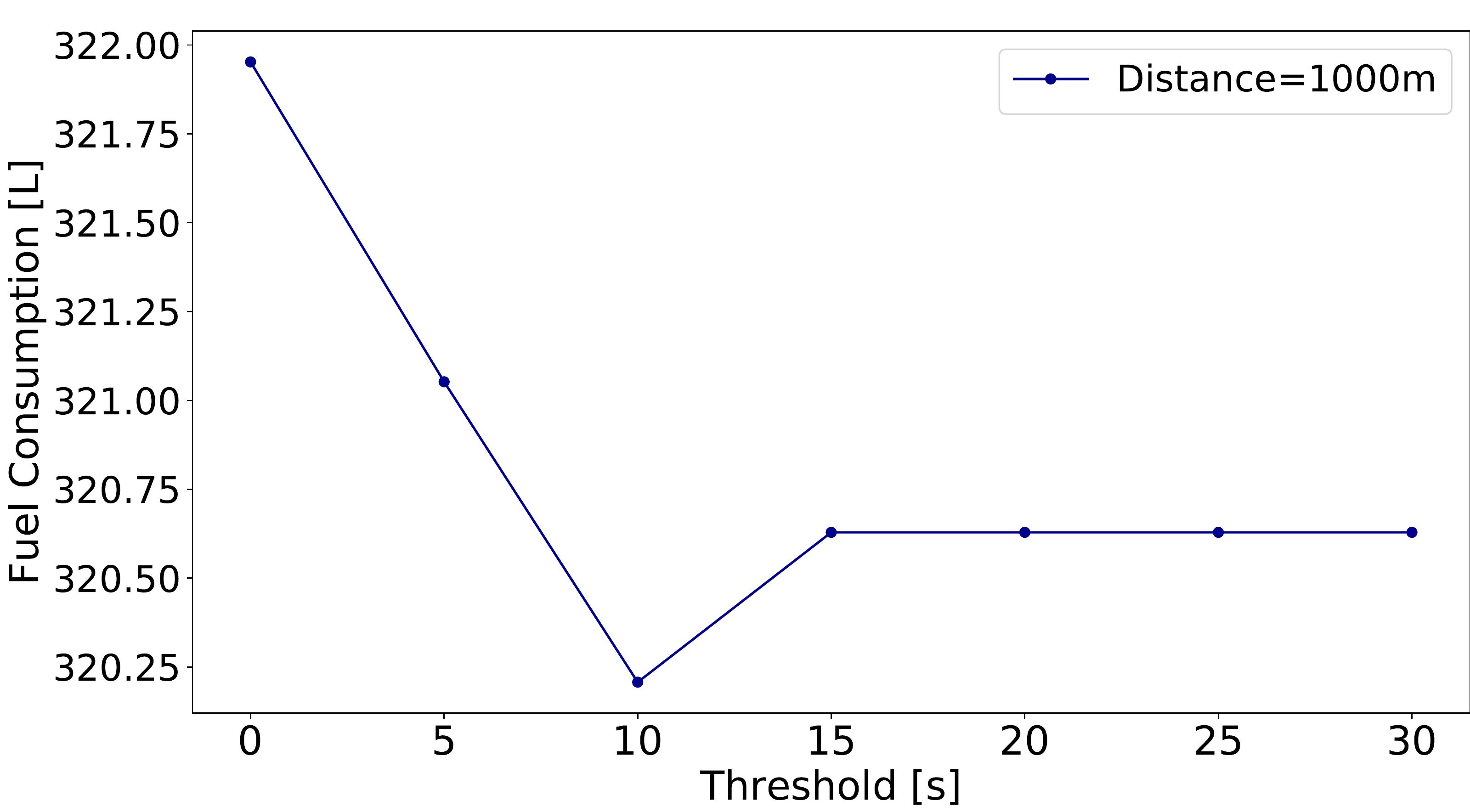}
  }
  
  %\subfigure
  {
  \includegraphics[width=0.5\textwidth]{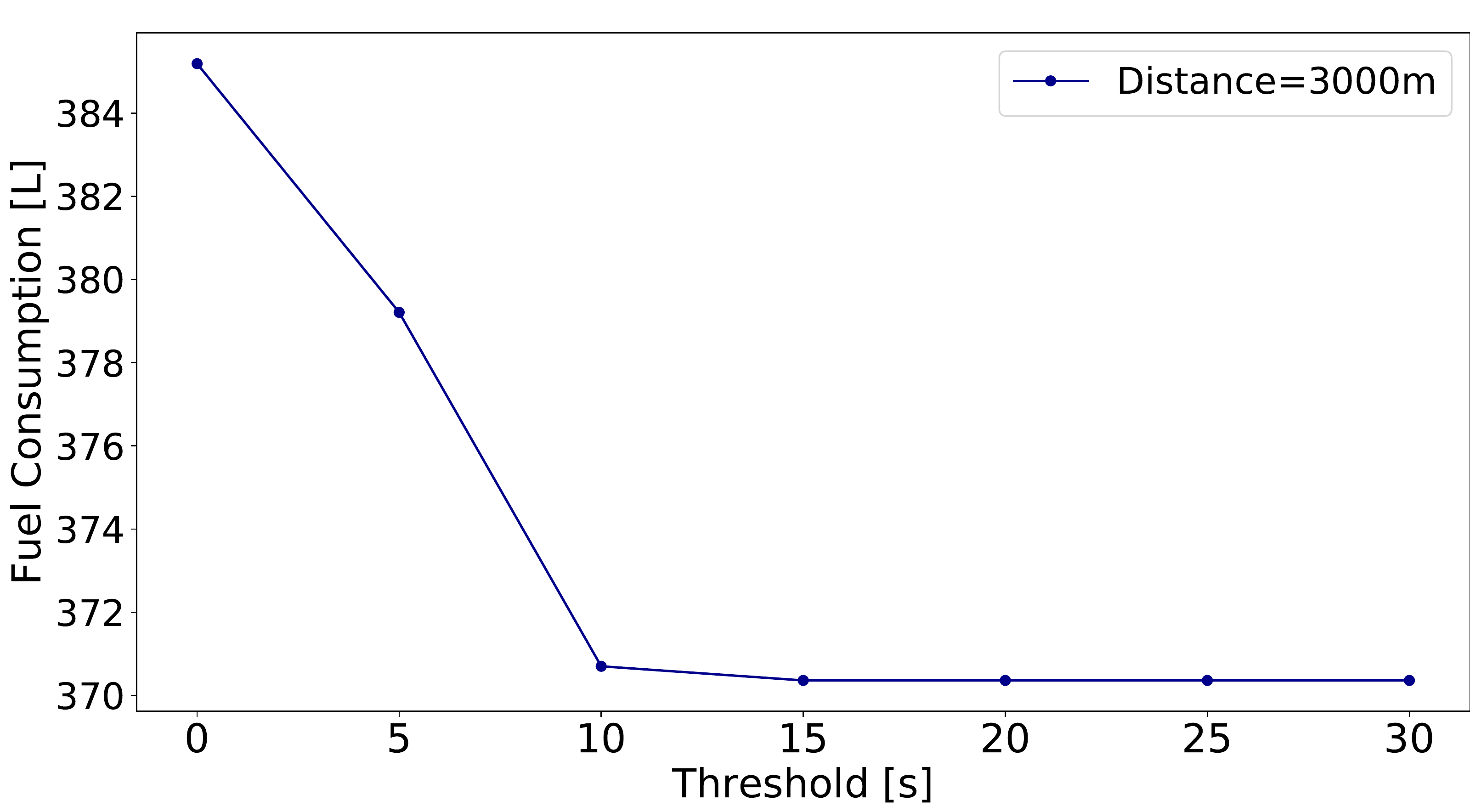}
  }
  
  %\subfigure
  {
  \includegraphics[width=0.5\textwidth]{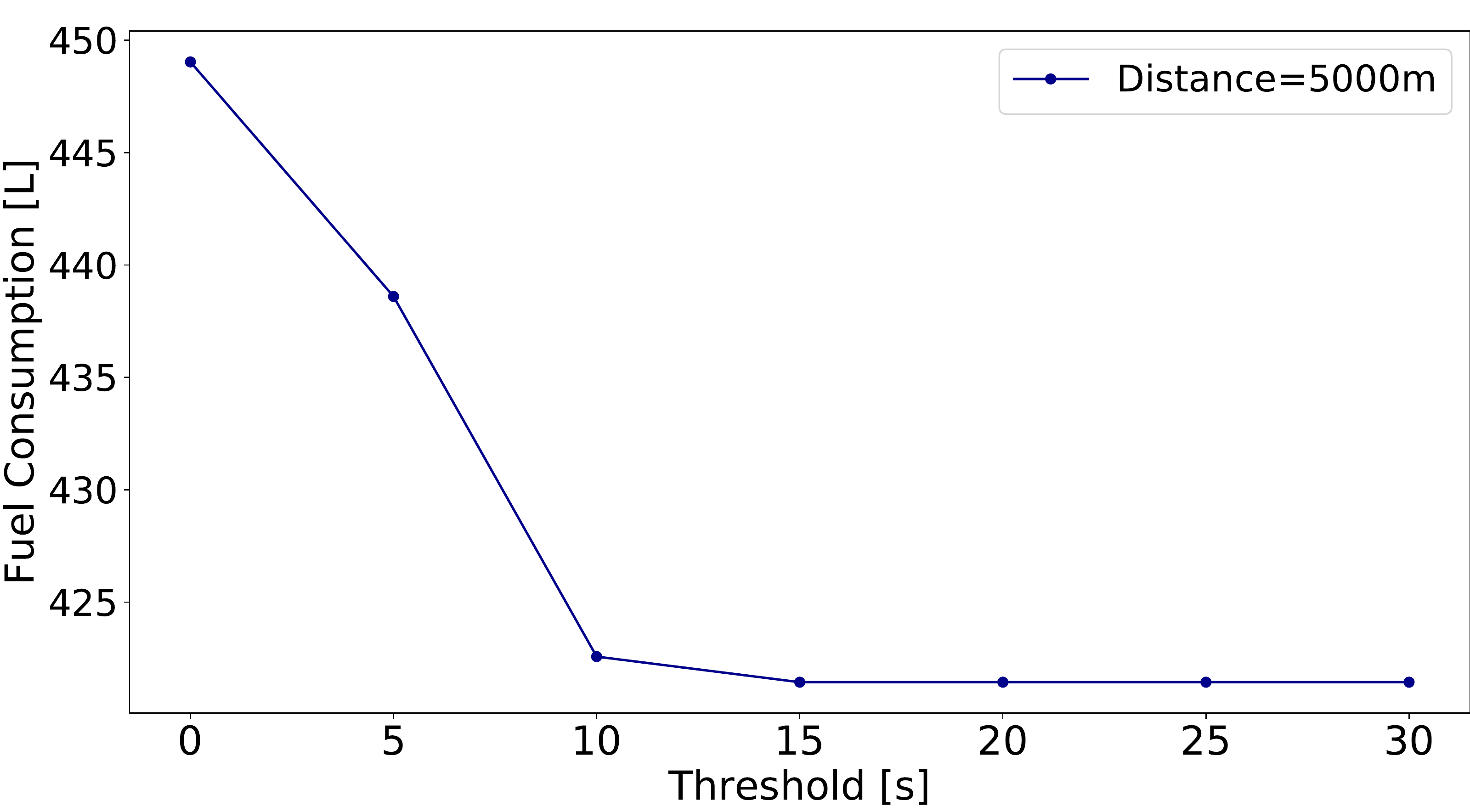}
  }
  \caption{Effect of cruising distance $D_2$.}\label{fig:D_2}
\end{figure}

In this section, we study the effect of cruising distance $D_2$ after the junction. Since the flow of heavy-duty vehicles are small, we assume connected heavy-duty vehicles are $0.1$ fraction of total demands in Table \ref{table:dataset}. We tested different distances of $D_2$: $500$ m, $1000$ m, $3000$ m, and $5000$ m, and the results are shown in Figure \ref{fig:D_2}.

We can see that when the distance of $D_2$ is small, the total fuel consumption would increase with the threshold. When the distance of $D_2$ is large, longer platoon would form easily, and total fuel consumption would decrease as we increase the threshold. These results are consistent with the implication in Equation \ref{Eq:final_fuel_consumption}. When the value of $D_2$ is small ($D_2 = 500$ m), the value $\Delta{F^k_2}$ approximates $0$, $\Delta{TC_k}$ would mainly consist of the acceleration fuel consumption $\Delta{F^k_1}$, which would result in the increased fuel consumption as we increase the threshold of forming platoons. When the curing distance $D_2$ is high ($D_2 = 3000$ m, $5000$ m), the vehicle would experience more fuel reduction due to the effect of $\eta$. When $D_2$ equals $1000$ m in Figure \ref{fig:D_2}, the curve first decreases and then increases as we increase the threshold, which indicates that there should exist an optimal threshold to obtain minimum fuel consumption. The optimal threshold can be calculated using the function in Equation \ref{Eq:final_fuel_consumption}.

\subsection{Homogeneous Threshold at All Junctions}
\begin{figure}[!ht]
  \centering
  %\subfigure
  {
  \includegraphics[width=0.7\textwidth]{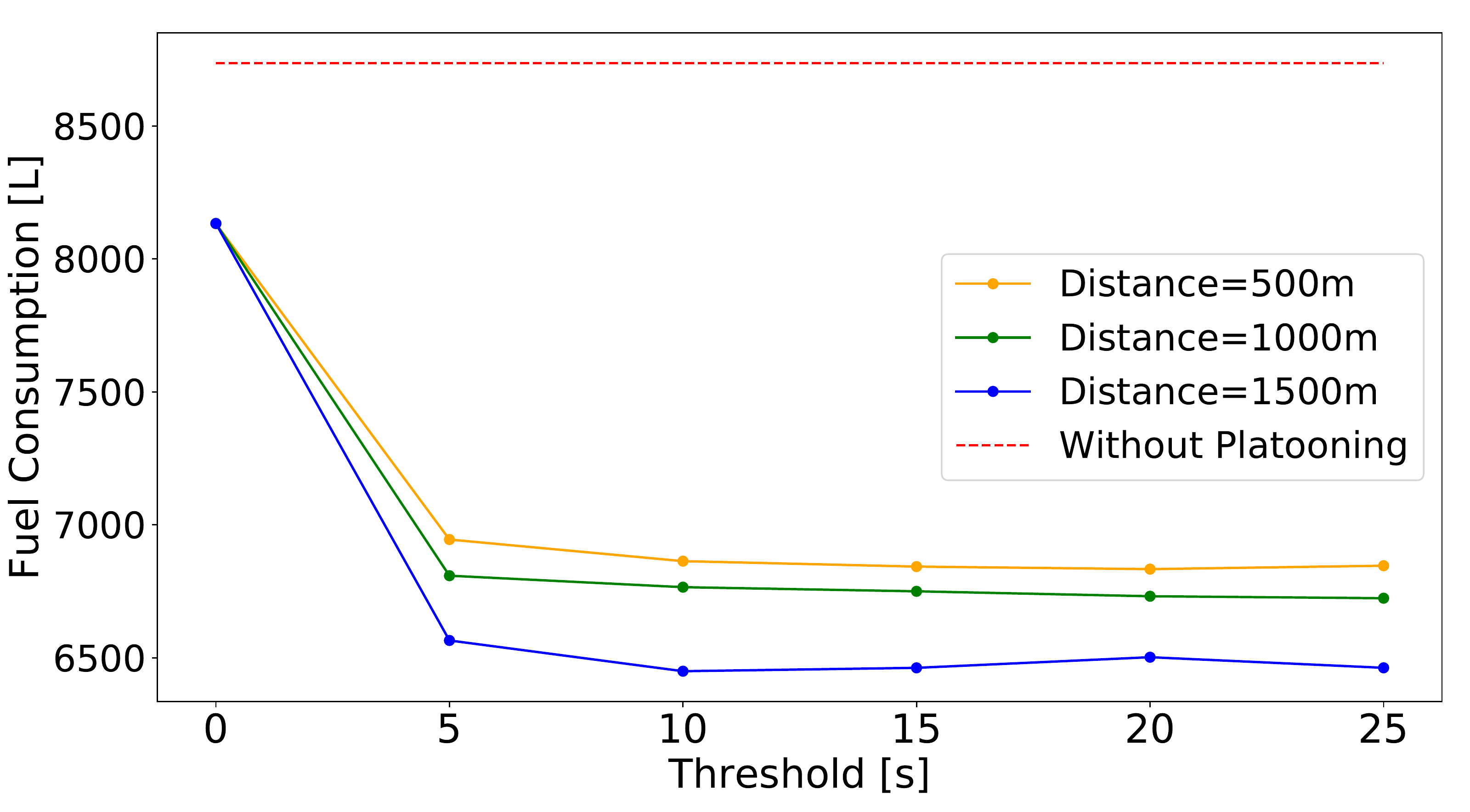}
  } \\(a) Effect of coordination distance $D_1$
  \label{fig:d_1}
  %\subfigure
  {
  \includegraphics[width=0.7\textwidth]{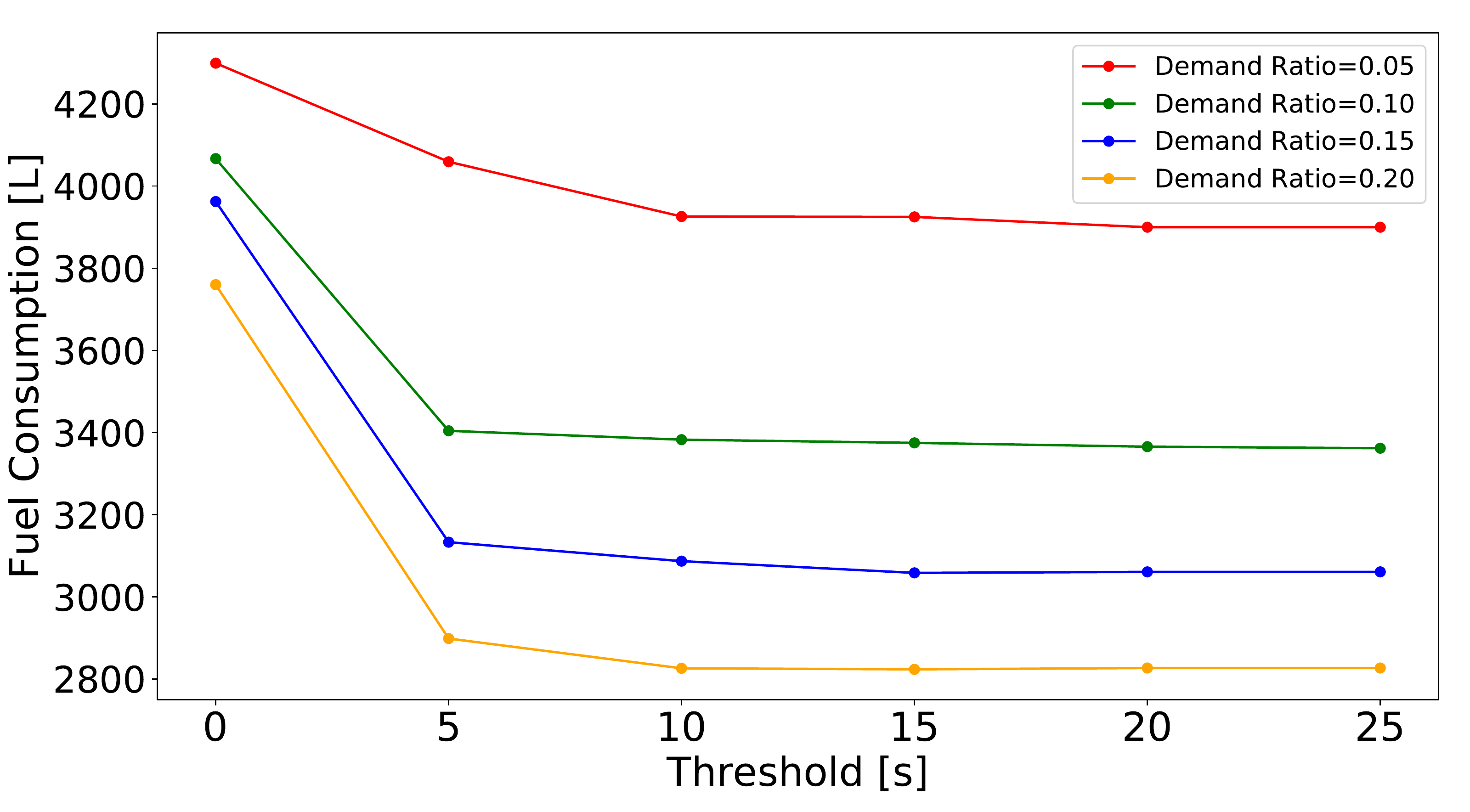}
  }\\(b) Effect of platoon ratio 
  \label{fig:ratio}
  \caption{Factors in homogeneous control}\label{fig:centralized_control}
\end{figure}

In this section we show the simulation results in homogeneous control at five junctions. Figure \ref{fig:centralized_control} shows the results using the same threshold for each junction. The results show that total fuel consumption would decrease as we increase the threshold to form more platoons. Total fuel consumption is significantly less than the value in the case without platoons. In networks, cruising distance $D_2$ is apparently very long, which weights the effect of $D_2$ over $D_1$

We also compare the effect of the placement of detectors. We have conducted experiments with three different $D_1$: $500$ m, $1000$ m, and $1500$ m. The results show that longer distance $D_1$ would decrease total fuel consumption, which is also consistent with Equation \ref{Eq:final_fuel_consumption}. Furthermore, longer distance of $D_1$ would also increase the chance of forming platoons.

In the previous sub-section, we investigate the effect of $D_2$ assuming the ratio of connected vehicles to be $0.1$. In this part, we use network simulation to show the effect of different ratios. Instinctively, more vehicles would lead to more fuel consumption. In order to compare the effect of connected vehicle ratio, we multiply the total fuel consumption by the ratio. In addition, we keep the value of $D_1$ to be constant and set the value to be $1000$ m. The results show that fuel consumption would decrease as we increase the ratio of connected vehicles, which shows that more platoons provide more benefits in the highway system.

\subsection{Heterogeneous Thresholds at Multiple Junctions}
In this section, we investigate heterogeneous control at different junctions. To simplify the simulation, we only study the coordination of two junctions. We chose junction 3 and junction 4 in Figure \ref{fig:networks} to implement the heterogeneous control. Each junction swipes the threshold from $5$ s to $25$ s, and the interval for threshold is $5$ s. During the simulation, thresholds in other three junctions remain the constant $20$ s.

\begin{figure}[!hbt]
  \centering
  \includegraphics[width=0.75\textwidth]{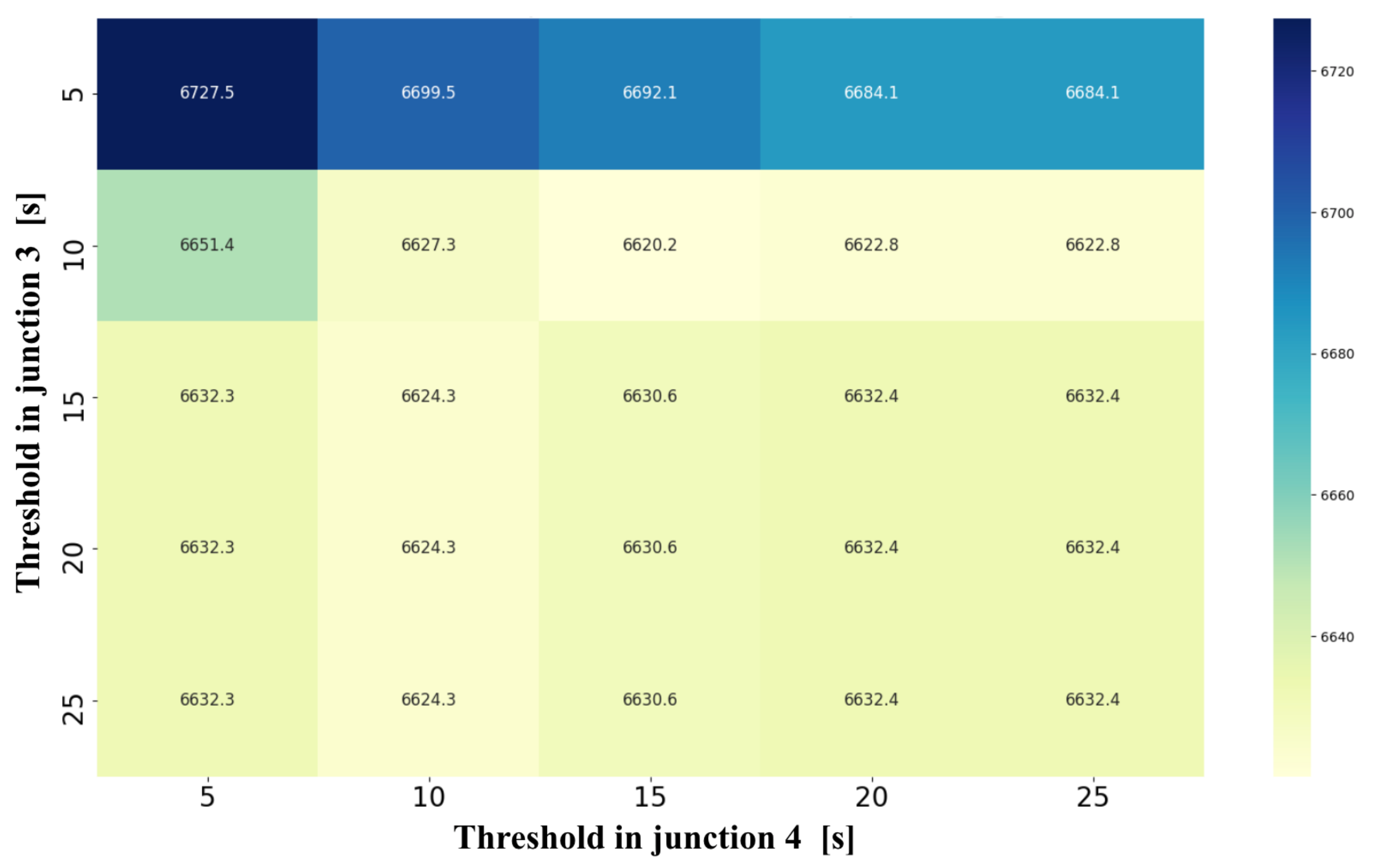}
  \caption{Coordinated control at different junctions} \label{fig:coordinated}
\end{figure}

Figure \ref{fig:coordinated} shows the heat map of coordinated control on the highway. The overall trend is still that larger threshold results in less fuel consumption. However, the system has the minimum value when the threshold in junction 3 is $10$ s, and the threshold in junction 4 is $15$ s, which indicates that the coordination system has minimum fuel consumption not necessarily at higher threshold. In Figure \ref{fig:coordinated}, when the threshold at junction 4 equals $15$ s, the fuel consumption would first decrease and then increase with the threshold at junction 3, which is consistent with the trend in Figure \ref{fig:D_2}.

In addition, the fuel consumption would decrease more if we increase the threshold at junction 3, which means that junction 3 has greater influence than junction 4. From the map in Figure \ref{fig:networks}, we can see that junction 3 locates on the left side of junction 4, and has longer cruising distance $D_2$, which would make the effect of junction 3 more significant.

%\begin{table}[!ht]
%  \begin{center}
%     \caption{Coordinated control at different %junctions}                       %\label{table:coordinated_control}
%    \begin{tabular}{ccccccc}
%    \hline
%    \multicolumn{2}{}{ \multirow{2}*{S_i} } & %\multicolumn{5}{c}{Threshold in junction 4 [s]} \\
%    \cline{3-7}
%   \multicolumn{2}{}{} &5 &10 &15 &20 &25          %    \\
%   \hline
% \multirow{5}*{\rotatebox{90}{Threshold in junction %3 [s]}} &5 &6727.5 &6699.5 &6692.1 &6684.1 &6684.1 %\\ 
% \cline{3-7}
%  &10 &6651.4 &6627.3 &6620.2 &6622.8 &6622.8  \\ 
% \cline{3-7}
%  &15 &6632.3 &6624.3 &6630.6 &6632.4 &6632.4  \\
% \cline{3-7}
%  &20 &6632.3 &6624.3 &6630.6 &6632.4 &6632.4 \\
% \cline{3-7}
%  &25 &6632.3 &6624.3 &6630.6 &6632.4 &6632.4 \\
% \hline
%\end{tabular}
%\end{center}
%\end{table}

%% file: sections/background.tex
\section{Impact of Background Traffic} \label{section:with_background_traffic}

In this section, we implement the simulation with connected vehicles and background vehicles. We also keep the ratio of connected vehicles to be 0.1, and then increase the ratio of background vehicles. Notice that the ratio here is referred to the percentage of the O-D demands in Table \ref{table:dataset}. As we add background vehicles into simulation, the computation efficiency would decrease. Then we reduce the simulation time to $1000$ s to implement this scenario.

Figure 6 shows the results of cases with background vehicles. When there is no background vehicle, the trend is still that fuel consumption would decrease with the threshold. When the ratio of background vehicles increases to $0.1$, the trend becomes unstable. As we add more background vehicles, the fuel consumption would increase finally, which indicates that background vehicles can change the behaviors of connected vehicles. When the ratio of background vehicles is 0.9, the incremental fuel consumption in $D_1$ is over the value in $D_2$.

\begin{figure}[!hbt]
  \centering \label{fig:background}
  %\subfigure
  {
  \includegraphics[width=0.48\textwidth]{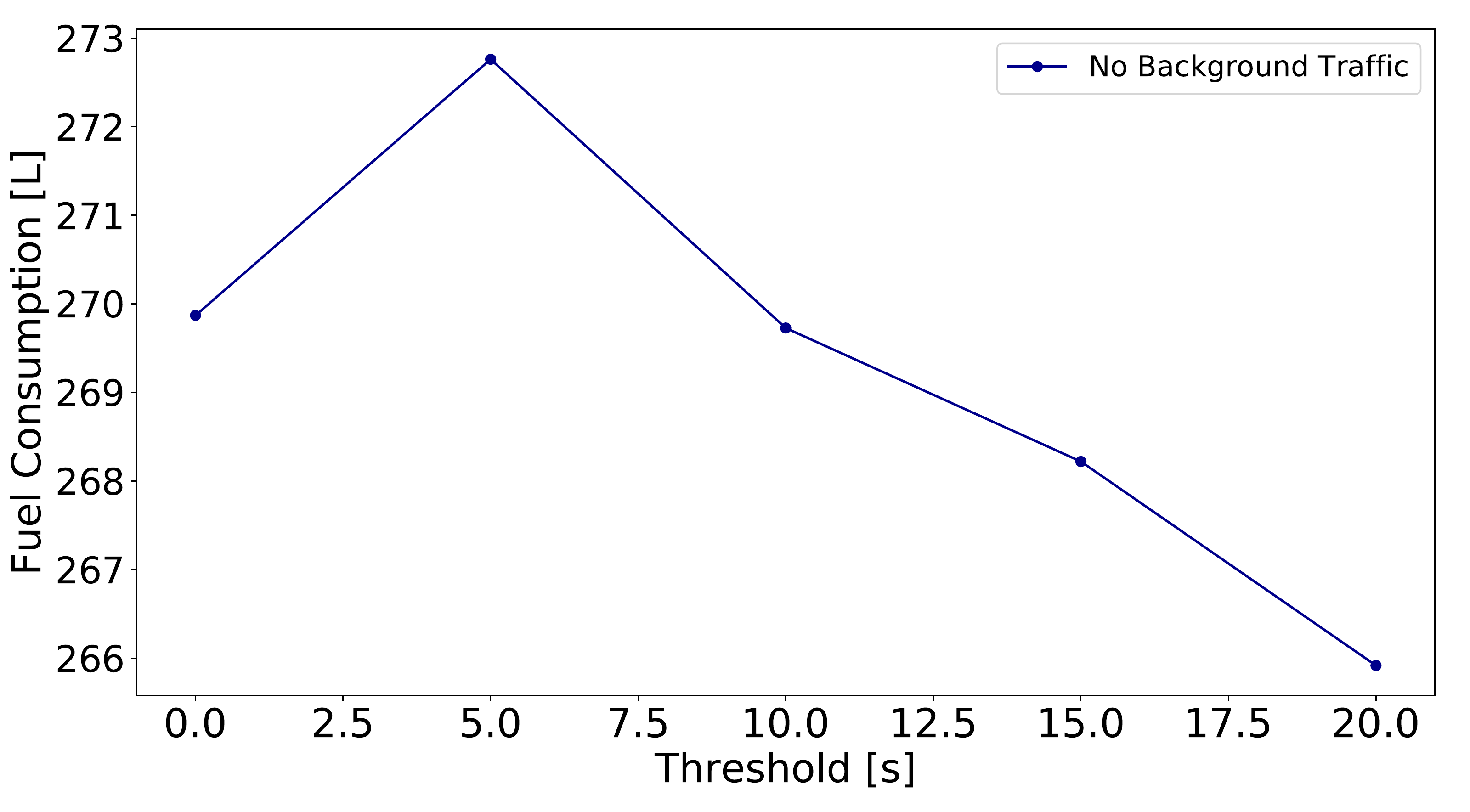}
  }
  %\subfigure
  {
  \includegraphics[width=0.48\textwidth]{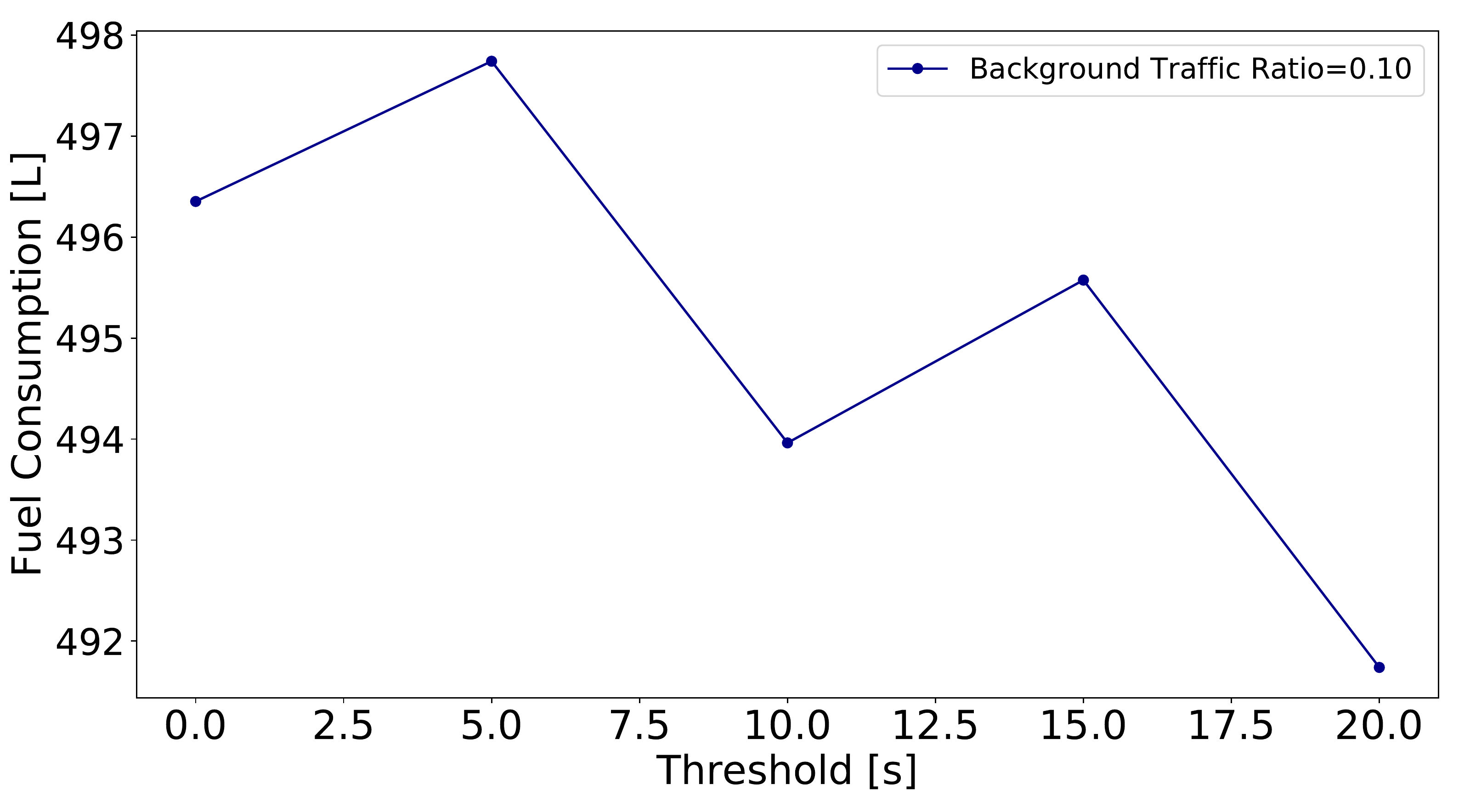}
  }
  %\subfigure
  {
  \includegraphics[width=0.48\textwidth]{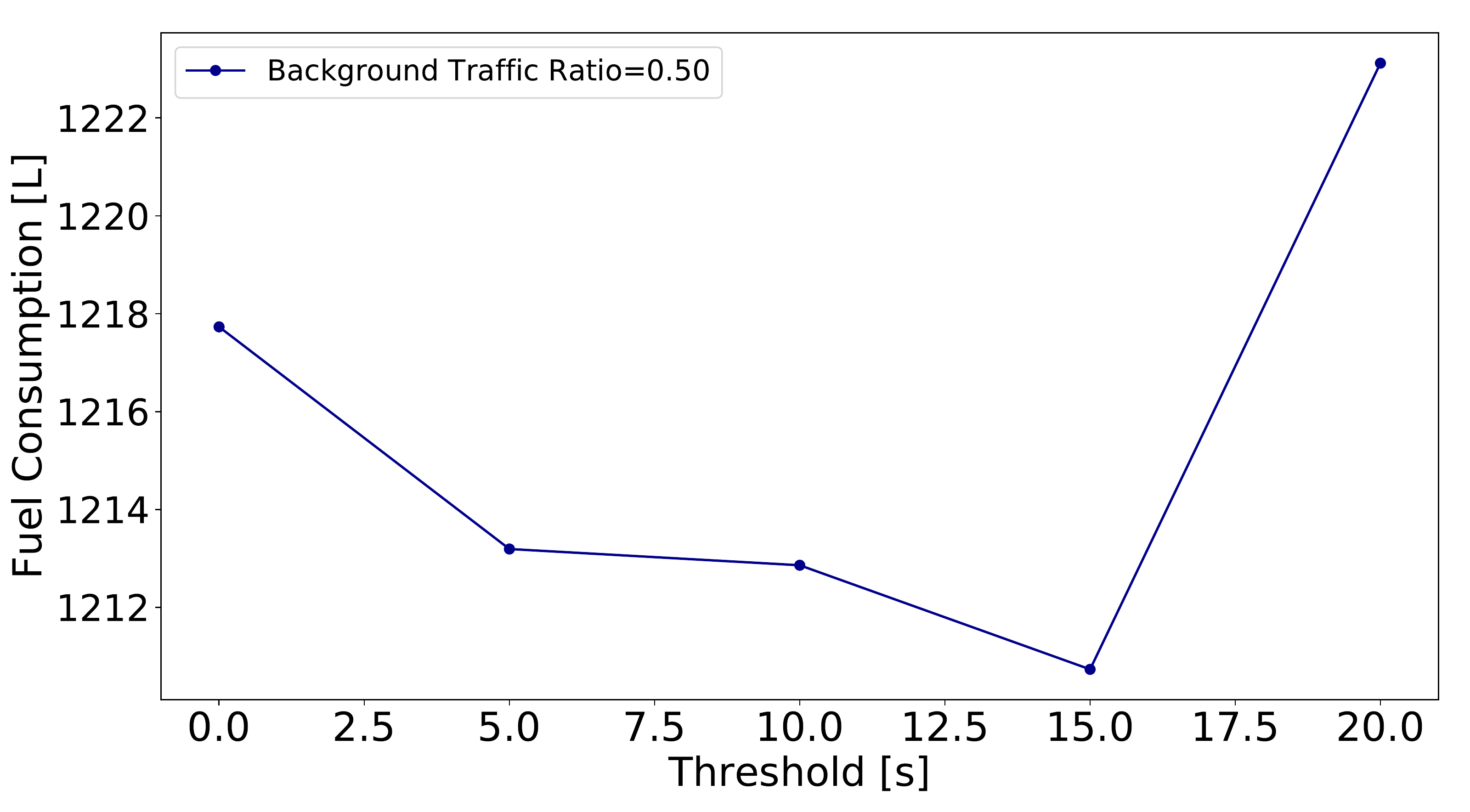}
  }
  %\subfigure
  {
  \includegraphics[width=0.48\textwidth]{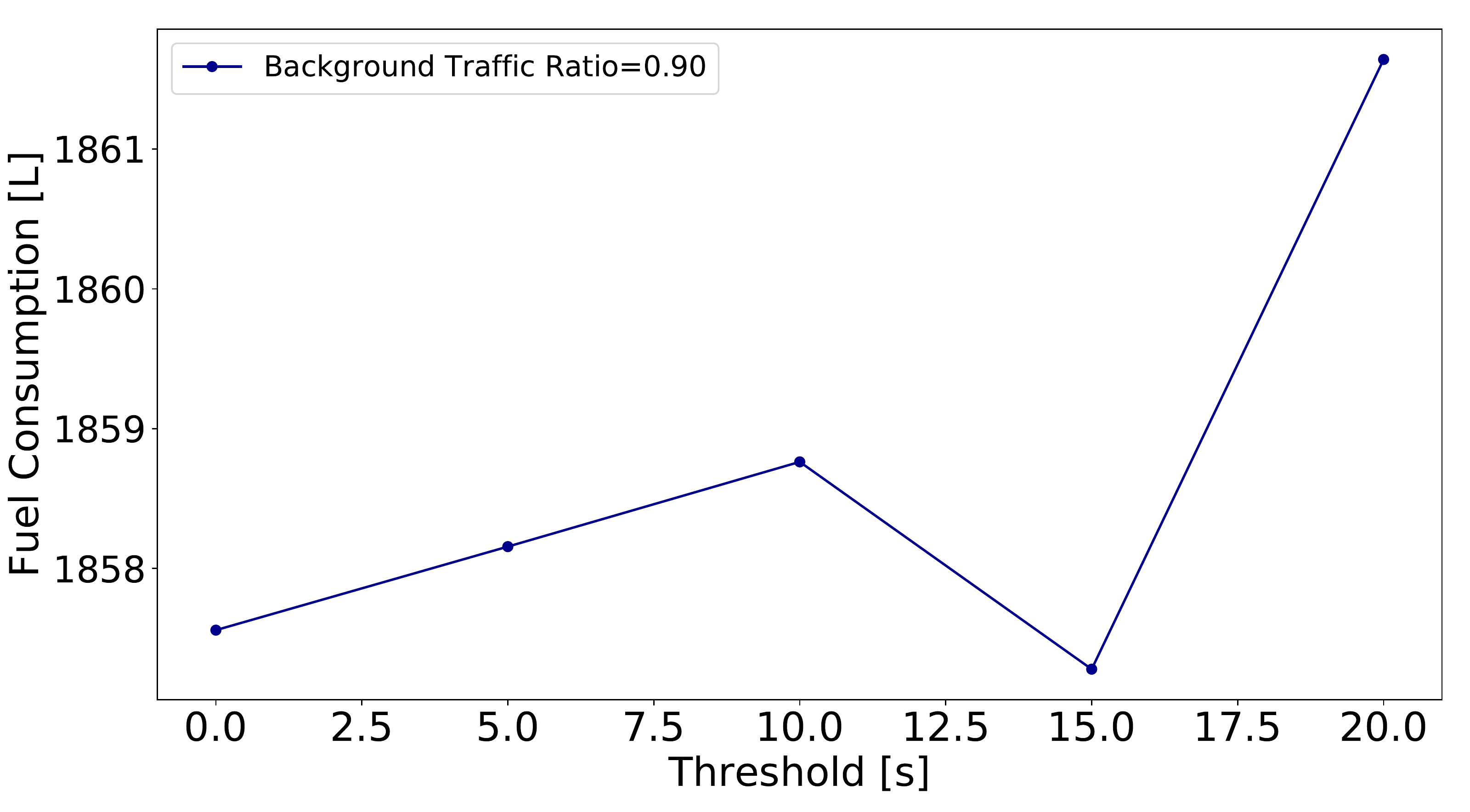}
  }
  \caption{Effect of background vehicles} 
\end{figure}

%% file: sections/conclusion.tex
\newpage
\section{Conclusions} \label{section:conclusion}

This article proposes a coordinated platooning system over a highway network. We present the hierarchical control system including upper level coordination and lower level platoon merging. The objective is to minimize total fuel consumption in coordination distance $D_1$ and cruising distance $D_2$. Real traffic data in PeMS is used to evaluate the performance of our proposed structure. We have conducted experiments in homogeneous and heterogeneous control in networks. The results show that there is a trade-off between incremental fuel in $D_1$ and reduced fuel in $D_2$. We also evaluate the scenarios with background vehicles, the results in mixed traffic show that background vehicle would influence the behaviors of connected vehicles.

This work can be extended in several directions. First, graph theory can be incorporated to determine platoon coordination in networks. Second, we can use multi-agent reinforcement learning to obtain optimal policy. Third, threshold sensitive to platoon size can be design to improve the flexibility of control policy.